\documentclass[twocolumn]{aastex63}
\newcommand{\kms}[0]{km s$^{-1}$}
\newcommand{\flx}[0]{erg s$^{-1}$ cm$^{-2}$}
\newcommand{\sbunit}[0]{erg s$^{-1}$ cm$^{-2}$ arcsec$^{-2}$}

\newcommand{\ha}[0]{H$\alpha$}
\newcommand{\delm}[0]{\textit{p}DSLM}
\newcommand{\delmfull}[0]{Dragonfly Spectral Line Mapper}
\newcommand{\objname}[0]{DF-E1}
\newcommand{\deblok}[0]{dB18}
\newcommand{\leroy}[0]{L15}
\newcommand{\yun}[0]{Y93}

\accepted{by ApJL}

\shorttitle{A TDG Forming Within M82's Northern HI Streamer}
\shortauthors{Pasha et al.}
\usepackage{float}
\graphicspath{{./}{figures/}}
\usepackage{CJK}

\begin{document}
\begin{CJK*}{UTF8}{gbsn}
%TC:ignore
\title{A Nascent Tidal Dwarf Galaxy Forming within the Northern HI Streamer of M82}

\correspondingauthor{Imad Pasha}
\email{imad.pasha@yale.edu}

\author[0000-0002-7075-9931]{Imad Pasha}
\altaffiliation{NSF Graduate Research Fellow}
\affiliation{Department of Astronomy, Yale University, 52 Hillhouse Ave., New Haven, CT 06511}

\author[0000-0002-2406-7344]{Deborah Lokhorst}
\affiliation{Department of Astronomy \& Astrophysics, University of Toronto, 50 St. George Street, Toronto, ON, M5S 3H4, Canada}
\affiliation{Dunlap Institute of Astronomy \& Astrophysics, University of Toronto, 50 St. George Street, Toronto, ON, M5S 3H4, Canada}

\author[0000-0002-8282-9888]{Pieter G. van Dokkum}
\affiliation{Department of Astronomy, Yale University, 52 Hillhouse Ave., New Haven, CT 06511}

\author[0000-0002-4175-3047]{Seery Chen}
\affiliation{Department of Astronomy \& Astrophysics, University of Toronto, 50 St. George Street, Toronto, ON, M5S 3H4, Canada}
\affiliation{Dunlap Institute of Astronomy \& Astrophysics, University of Toronto, 50 St. George Street, Toronto, ON, M5S 3H4, Canada}

\author[0000-0002-4542-921X]{Roberto Abraham}
\affiliation{Department of Astronomy \& Astrophysics, University of Toronto, 50 St. George Street, Toronto, ON, M5S 3H4, Canada}
\affiliation{Dunlap Institute of Astronomy \& Astrophysics, University of Toronto, 50 St. George Street, Toronto, ON, M5S 3H4, Canada}

\author[0000-0003-4970-2874]{Johnny Greco}
\affiliation{Center for Cosmology and AstroParticle Physics (CCAPP), The Ohio State University, Columbus, OH 43210, USA}

\author[0000-0002-1841-2252]{Shany Danieli}
\affiliation{Department of Astronomy, Yale University, 52 Hillhouse Ave., New Haven, CT 06511}

\author[0000-0001-8367-6265]{Tim Miller}
\affiliation{Department of Astronomy, Yale University, 52 Hillhouse Ave., New Haven, CT 06511}

\author[0000-0002-8543-2528]{Erin Lippitt}
\affiliation{Department of Astronomy, Yale University, 52 Hillhouse Ave., New Haven, CT 06511}

\author[0000-0002-5283-933X]{Ava Polzin}
\affiliation{Department of Astronomy, Yale University, 52 Hillhouse Ave., New Haven, CT 06511}

\author[0000-0002-5120-1684]{Zili Shen}
\affiliation{Department of Astronomy, Yale University, 52 Hillhouse Ave., New Haven, CT 06511}

\author[0000-0002-7743-2501]{Michael A. Keim}
\affiliation{Department of Astronomy, Yale University, 52 Hillhouse Ave., New Haven, CT 06511}
\author[0000-0002-7490-5991]{Qing Liu (刘青)}
\affiliation{Department of Astronomy \& Astrophysics, University of Toronto, 50 St. George Street, Toronto, ON, M5S 3H4, Canada}
\affiliation{Dunlap Institute of Astronomy \& Astrophysics, University of Toronto, 50 St. George Street, Toronto, ON, M5S 3H4, Canada}

\author[0000-0001-9467-7298]{Allison Merritt}
\affiliation{Max-Planck-Institut f{\"u}r Astronomie, K{\"o}nigstuhl 17, D-69117 Heidelberg, Germany}

\author[0000-0001-5310-4186]{Jielai Zhang}
\affiliation{ Centre for Astrophysics and Supercomputing, Swinburne University of Technology, Mail Number H29, PO Box 218, 31122 Hawthorn, VIC, Australia}
\affiliation{ARC Centre of Excellence for Gravitational Wave Discovery (OzGrav), Hawthorn, 3122, Australia}

\begin{abstract}
We identify a $\sim$600 pc-wide region of active star formation located within a tidal streamer of M82 via \ha{} emission ($F_{H\alpha}\sim 6.5\times10^{-14}$ erg s$^{-1}$ cm$^{-2}$), using a pathfinder instrument based on the Dragonfly Telephoto Array. The object is kinematically decoupled from the disk of M82 as confirmed via Keck / LRIS spectroscopy, and is spatially and kinematically coincident with an overdensity of HI and molecular hydrogen within the ``northern HI streamer" induced by the passage of M81 several hundred Myr ago. From HI data, we estimate that $\sim5\times10^7$ M$_{\odot}$ of gas is present in the specific overdensity coincident with the \ha{} source. The object's derived metallicity (12+$\log(O/H)\simeq 8.6$), position within a gas-rich tidal feature, and morphology (600 pc diameter with multiple star forming clumps), indicate that it is likely a tidal dwarf galaxy in the earliest stages of formation.
\end{abstract}


\section{Introduction} \label{sec:intro}
M82 is among the most extensively-studied galaxies, with decades of high resolution observations having been facilitated by the M81 group's close proximity \citep[D $\simeq$ 3.63 Mpc;][]{Karachentsev:2004}. M82 is considered the ``prototypical" starbursting galaxy, featuring a bipolar superwind outflow that extends kiloparsecs along the minor axis of the galaxy. Many of M82's features, including a disturbed disk and the starburst itself, are thought to have been triggered by the recent passage of M81 $\sim$200$-$300 Myr ago \citep[e.g.,][]{Cottrell:1977}. Studies of the neutral and molecular hydrogen in M82 have found evidence of tidally induced structures, including disruption of the gas from the disk of M82 \citep[e.g.,][]{Sofue:1998} and a significant amount of gas located in the field and in tidal bridges between M81, M82 and NGC 3077 \citep[e.g.,][]{Yun:1994}. 

This environment makes the M81 group conducive to the formation of Tidal Dwarf Galaxies (TDGs), which form in the tidal arms and bridges induced by galaxy encounters. Gravitational instabilities develop in the absence of the rotational support that existed in the disk, and the Jeans mass rises due to the additional dispersion created by the heating of the material during the encounter \citep[e.g.][]{Duc:2000,Struck:2005}. In contrast to typical low mass dwarf galaxies, TDGs have little to no dark matter, and have metallicities commensurate with their progenitor spirals rather than the range expected from the dwarf mass-metallicity relation \citep[e.g.,][]{Duc:2004,Bournaud:2006}. The creation rate of TDGs and their survivability rates are poorly constrained, but have potential impacts on the interpretation of satellite population statistics in a $\Lambda$CDM framework.

Several of the dwarf galaxies in the M81 group are thought to be TDGs, including Holmberg IX, BK3N, Arp-loop (A0952+69), and Garland \citep[e.g.,][]{Makarova:2002,Weisz:2008,Sabbi:2008,deMello:2008}; however, none of these candidate TDGs or TDG progenitors are located in the vicinity of M82, instead clustering near M81 and NGC 3077. Indeed, no regions of active star formation have been reported in and around M82 outside the galactic disk.

We have used a novel telescope built on the framework of the Dragonfly Telephoto Array \citep[][]{Abraham:2014}, the 3-lens narrowband \textit{pathfinder} version of the forthcoming \delmfull{} (\delm{}), to image low surface brightness emission in the  M81 group in \ha{} and [NII]. We selected the M81 group to vet the narrowband design because deep prior \ha{} observations exist, providing a test of the depths being reached by the instrument. Our final \ha{} images of the field reach surface brightness depths of $\sim5\times 10^{-19}$ \sbunit{}, and uncover several new features in the gas around M82 that will be described in forthcoming work. Details of the \delm{} instrument are provided in \cite{Lokhorst:2019,Lokhorst:2020}.

In this Letter, we present the analysis of an \ha{} object near M82 that is likely a TDG in formation, using the first science data obtained with this instrument. We have designated this source \objname{} (``Sapling"), as it is the first emission source cataloged by Dragonfly.  

\begin{figure*}
    \centering
    \includegraphics[width=\textwidth]{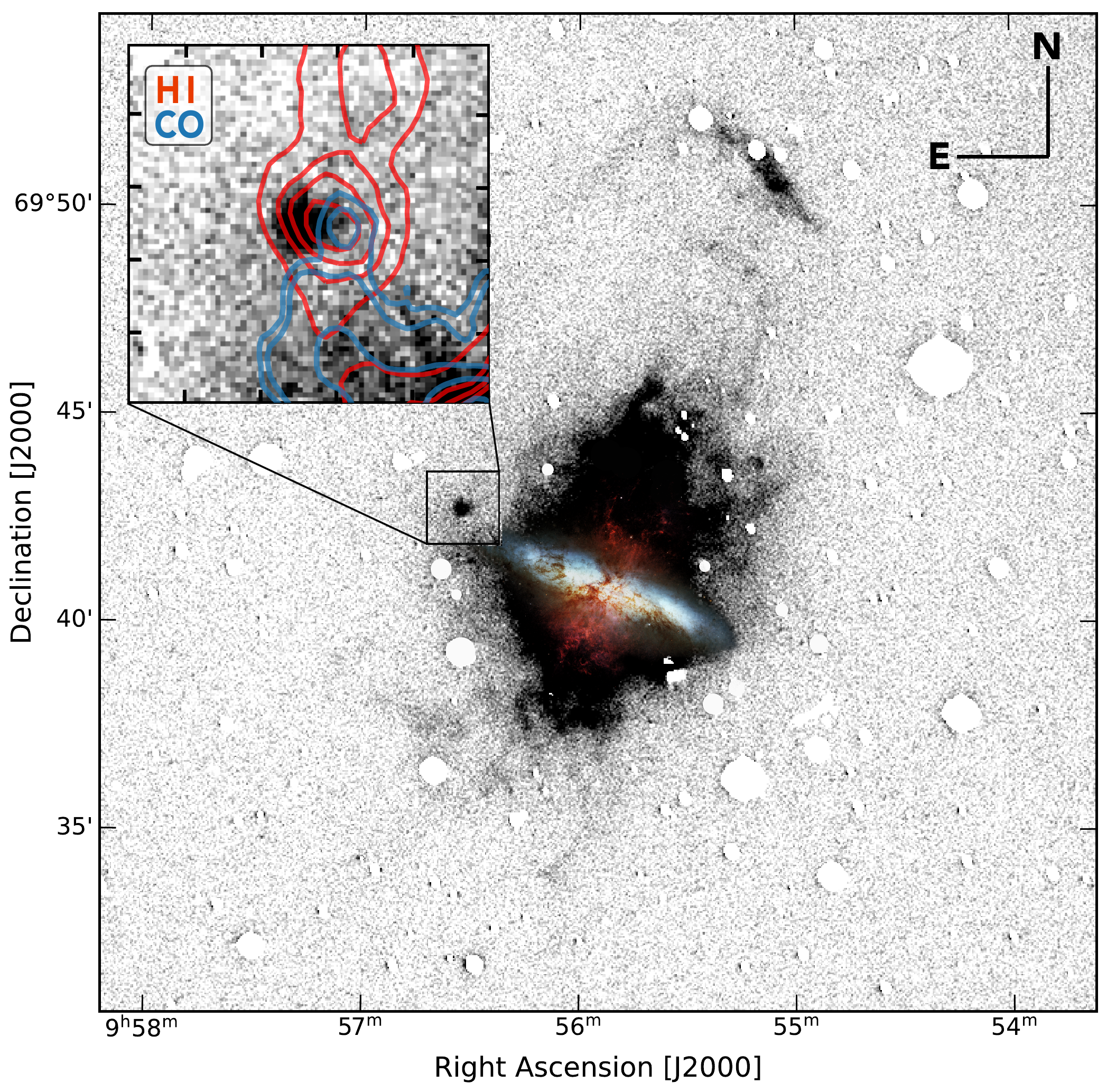}
    \caption{\delm{} continuum-subtracted image of M82 in \ha{}, with foreground stars masked. The \textit{HST} composite image of M82 from \cite{Mutchler:2007} has been overlaid for a sense of scale. An inset panel shows a 2.5$'$ $\times$ 2.5$'$ square cutout around \objname{}, with red contours showing HI from \cite{deblok:2018} (\deblok{} hereafter) and blue contours showing CO from \cite{Leroy:2015} (\leroy{} hereafter) in the region overlaid. The strongest overdensity of the northern HI tidal arm spatially overlaps with \objname{}, with the peak of the HI emission falling within 15$''$ of the peak of the \ha{} source. Similarly, an overdensity of CO is present, consistent with the HI peak. }
    \label{M82_crop}
\end{figure*}

\section{Narrowband Imaging}
\subsection{Observations}
 The \delm{} builds on the telephoto array concept of Dragonfly by placing ultra narrowband interference filters ($\Delta\lambda=0.8-3$ nm) in front of each lens of the array, targeting \ha{} and [NII] via tilting of the filters, and [OIII].
 
 The M81 group was observed with the \delm{} over the course of several weeks during the spring of 2020 at both \ha{} and [NII] tilts. Observations were carried out semi-autonomously, with a dither pattern of 15$'$ between exposures. Individual frames had exposure times of 1800 seconds. During data reduction, low quality frames were automatically identified and removed. Frames ultimately used in this analysis were also checked by eye to ensure quality; some frames were removed due to light contamination from sources in the dome or were cropped due to significant amplifier glow in corners of the frame.

The final images presented here comprise 190 frames in \ha{} (taken across the three lenses in the pathfinder array), for a ``single-lens" total exposure time of 95 hours, and 89 frames in [NII] for an total exposure time of 44.5 hours. Individual images were flatfielded and dark current subtracted using calibration frames taken nearest each observation temporally. 

\subsection{Continuum Subtraction and Flux Calibration}
Continuum subtraction was performed using exposures taken with the Dragonfly Telephoto Array in the Sloan-$r$ band. These images have similar pixel scales and point spread functions to the images taken with the \delm{} setup. An effective scaling factor of 11.3 was determined between the continuum image and the narrow band images, using a combination of aperture photometry on stars in the field as well as apertures placed on the disk of M82. We used the \texttt{SourceExtractor} package \citep{Bertin:1996} to identify and mask any point-like sources in images prior to continuum subtraction.

Published fluxes of HII regions in M81 were used to directly calibrate our image counts; while not shown in the cutouts presented here, M81 was in the imaging field of view. We selected two independent datasets to use, one from \cite{Lin:2003}, the other from \cite{Patterson:2012}, which present spectrally-calibrated \ha{}+[NII] fluxes and \ha{} luminosities, respectively. We defined one calibration by combining our [NII] and \ha{} images and comparing to \cite{Lin:2003}, and another by comparing to the fluxes implied by the \cite{Patterson:2012} luminosities to our \ha{} image alone.

\begin{deluxetable}{lr}[b]
\tablehead{
\colhead{Source Property} & \colhead{\hspace{4.5cm}Value\hspace{.5cm}}
}
\caption{Properties of \objname{}}
\startdata 
R.A. & 09 56 32.184\\
DEC & +69 42 45.453 \\
$v_{\rm helio}$ & 276.7 ($\pm$ 25) \kms{}\\
$v_{\rm helio}-v_{\rm M82}$ & 73.7 ($\pm$ 25) \kms{}\\
Diameter\tablenotemark{\footnotesize a} (arcsec) & 34.6$''$ \\
Diameter\tablenotemark{\footnotesize b} (pc) &  608 pc \\
\ha{} Flux\tablenotemark{\footnotesize c} & (6.5$\pm$0.9)$\times$10$^{-14}$ \flx{} \\
{[}NII{]} Flux\tablenotemark{\footnotesize c} & (2.9$\pm$0.4)$\times$10$^{-14}$ \flx{} \\
{[}NII{]}/\ha{} (\delm{}) & 0.38 $\pm$ 0.06\\
{[}NII{]}/\ha{} (LRIS) & 0.31 $\pm$ 0.015\\
{[}OIII{]}/$H\beta$ (LRIS) & 0.34 $\pm$ 0.07 \\
12+$\log(O/H)$ (via $O3N2$) & 8.67 $\pm$ 0.14 \\
L(\ha{})\tablenotemark{\footnotesize c} & (1.03$\pm$0.14)$\times$10$^{38}$ erg s$^{-1}$\\
log SFR(\ha{}) & $-$2.39 $\pm$ 0.03\\
FUV Flux Density & (2.73$\pm$1.1)$\times$10$^{-14}$ \flx{} \AA$^{-1}$ \\
log SFR(FUV)\tablenotemark{\footnotesize c}  & $-$2.317 \\
HI Mass & $\sim$5$\times$10$^7$ M$_{\odot}$\\
\enddata
\tablenotetext{a}{\footnotesize Defined as the mean FWHM of a Gaussian fit to the Dragonfly image.}
\tablenotetext{b}{\footnotesize Assuming a distance of 3.63 Mpc.}
\tablenotetext{c}{\footnotesize Not corrected for dust attenuation.}
\end{deluxetable}

For each calibration, we computed best-fit relations in log space using a first degree polynomial in an iterative sigma-clipping routine that eliminates outliers of $>3\sigma$ in each iteration. Both calibrations were consistent with each other, with no observed systematic offset. Due to the greater number of HII regions included and marginally tighter scatter, we have adopted the best fit relation to the \cite{Lin:2003} observations in this work, which has a derived calibration of 
\begin{equation}
    \log(F)\; \rm [erg\; s^{-1}\; cm^{-2}] = 0.99\log(cts) - 16.801,
\end{equation}
where log is base 10 (adopted throughout). Based on the scatter in the fit relation, we estimate the calibration uncertainty is $\sim$5\% in flux. Additional details regarding the instrument, reduction and calibration are provided in a companion paper (Lokhorst et al., submitted).

\section{An \ha{} Object Near M82}

A summary of measurements presented in the following sections is provided in Table 1.

\subsection{The \ha{} feature seen with the \delm{}}
An \ha{} image from \delm{} of M82 is presented in Figure \ref{M82_crop}. \objname{} is $\sim$4.3 kpc from the galactic center and 493 pc above the midplane \citep[$\sim$3 $h_z$;][]{Lim:2013}. While several \ha{} maps from previous studies \citep[e.g.,][]{Lehnert:1999,Hoopes:2005,Karachentsev:2007} show evidence for \ha{} emission overdensities at \objname{}'s location, the detections were not discussed in those works. No study to our knowledge has specifically referenced this \ha{} object.

\begin{figure*}[t]
    \centering
    \includegraphics[width=\textwidth]{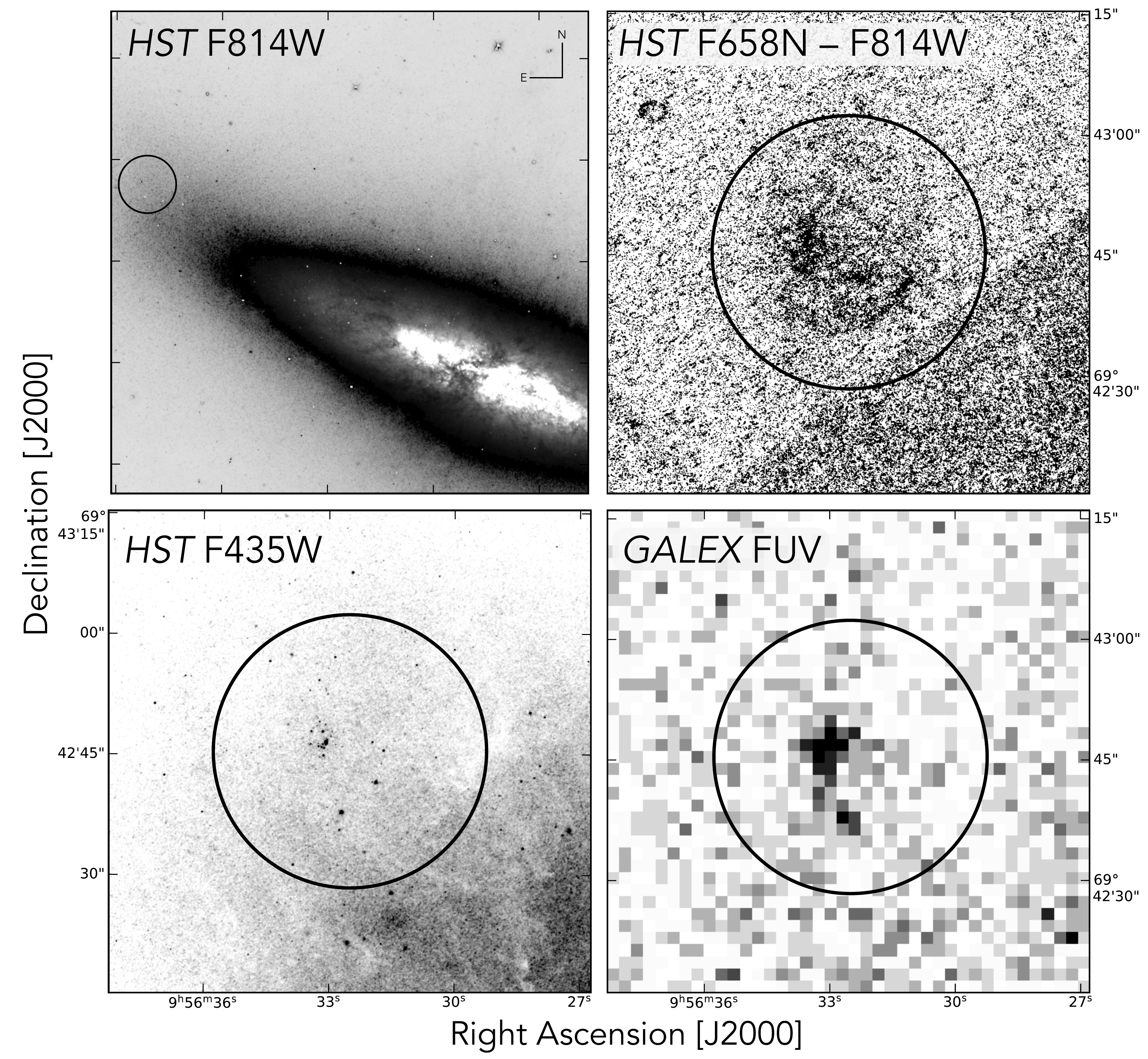}
    \caption{\objname{} in \textit{HST} F814W (top left), continuum-subtracted F658N (top right), F435W (bottom left), and \textit{GALEX} FUV (bottom right). The F814W panel serves to contextualize \objname{}'s position and size with respect to M82; note, the inner part of M82 is shown in an inverse color scaling for visual clarity. The other three panels demonstrate a strong correspondence between the \ha{} emission morphology of \objname{}, resolved point sources in \textit{HST} F435W, and FUV emission.}
    \label{hst_comparison}
\end{figure*}

\objname{} has a well defined peak in the Dragonfly data, and exhibits lobes of elongation in both the north-south and east-west directions, resembling an ``L-shaped" blob. We measure the size of \objname{} from the \delm{} imaging by fitting Gaussians to the flux profile of the source along both the north-south and east-west position angles, passing through the peak flux location. The average of the two fits was kept. The two fits were nearly identical, with a full width half maximum (FWHM) value of 34.6$''$ and 95\% width of 58.8$''$, corresponding to linear scales of 608 pc and 1034 pc, respectively. 

\begin{figure*}[t]
    \centering
    \includegraphics[width=\textwidth]{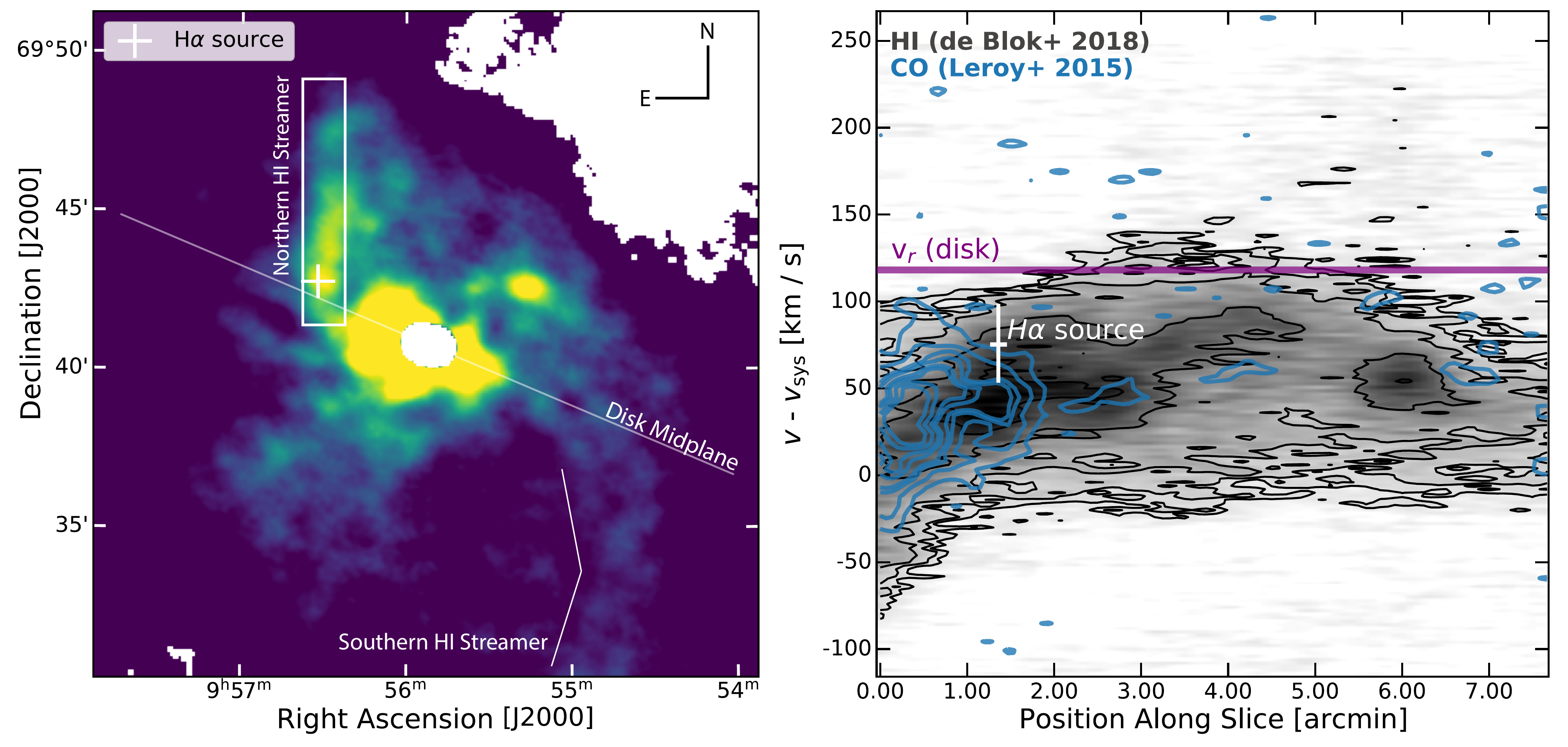}
    \caption{HI Moment 0 map from \deblok{} (left), with tidal features and the disk midplane labeled. In the right panel, we extract a position-velocity diagram along the slice indicated by the white aperture in the left panel from both the HI and CO data. The location of \objname{} in position-velocity space is indicated by a plus, demonstrating its velocity correspondence with the tidal feature.}
    \label{HI_comparison}
\end{figure*}

The flux of the source in the aperture defined by the FWHM is $\sim$6.5$\times$10$^{-14}$ \flx{}. We did not explicitly correct for a background contribution to the flux from M82, but note that in the region around but not including the source, the \ha{} background is $\sim$10\% of the measured flux, which we adopt into our uncertainty. We measured the [NII] flux similarly, obtaining $F_{[NII]}=$2.9$\times$10$^{-14}$ \flx{}, corresponding to an [NII]/\ha{} ratio of 0.38$\pm$0.06, in line with typical values for star forming regions, and in general lower than the disk of M82, which is known to have elevated [NII]/\ha{} ratios due to turbulence and shocks \cite[e.g.,][]{Shopbell:1998}.

The corresponding dust-uncorrected \ha{} luminosity of the source is 1.03$\times$10$^{38}$ erg s$^{-1}$. To estimate a star formation rate from this value, we used the calibration from \cite{Lee:2009}: 

\begin{equation}
    \log SFR\; [M_{\odot}\; \textrm{yr}^{-1} ] = 0.62 \log \left(7.9\times10^{42} L_{H\alpha}\right) - 0.47,
\end{equation}
where $L_{H\alpha}$ is in erg s$^{-1}$. We selected this calibration as it is (a) tuned to local dwarf galaxies with \ha{} luminosities in the range of 10$^{37}$ to 2.5$\times$10$^{39}$ erg s$^{-1}$, matching our source and (b) is calibrated using the extinction-corrected FUV-\ha{} relation to be intrinsically dust-correcting. Using this calibration, we obtained an SFR of 0.004 $M_{\odot}$ yr$^{-1}$ (log SFR $= -2.39$). 

\subsection{Far-UV and Optical Imaging}
If \objname{} is indeed forming stars, then we expect cospatial FUV emission. In Figure \ref{hst_comparison} we present \textit{Galaxy Evolution Explorer} (\textit{GALEX}) imaging from \cite{Hoopes:2005} of M82, as well as F435W, F814W, and F658N imaging from \textit{HST} \citep[][]{Mutchler:2007}. We identify several regions of concentrated FUV emission within the extent of \objname{}, indicating the existence of young O/B stars. That this emission is not a background source is confirmed by the F435W imaging, which resolves the FUV clumps into several regions containing one or more bright point sources, and at least one clustering of stars, which we note are found to form efficiently in TDGs \citep{Fensch:2019}. These sources furthermore, align with the ionized gas in the region.

Because TDGs form in tidal arms where dust content is expected to be minimal compared to the host galaxy, FUV emission provides a useful direct probe of ongoing star formation. Leveraging this expected consistency, we use the \ha{}-derived SFR to infer the expected ionizing FUV flux, and compare this with the \textit{GALEX} data. We convert the \ha{}-derived SFR via \cite{Kennicutt:1998}:
\begin{equation}
    SFR\; [M_{\odot}\; \textrm{yr}^{-1}] = 1.4\times10^{-28}\; L_{FUV}\; [\rm erg\; s^{-1}\; Hz^{-1}],
\end{equation}
which predicts a FUV flux of 1.83$\times$10$^{-26}$ erg s$^{-1}$ cm$^{-2}$ Hz$^{-1}$. We measure the FUV flux directly in an 11$''$ aperture, as the emission is more concentrated than the \ha{}. Given the extremely low UV background in \textit{GALEX} FUV, we assume Poisson uncertainties on the counts. This produces a FUV flux of (2.2$\pm$0.5)$\times$10$^{-26}$ erg s$^{-1}$ cm$^{-2}$ Hz$^{-1}$, consistent with the predicted FUV value. In the absence of resolved \textit{HST} imaging, it is clear from Figure \ref{hst_comparison} that UV emission is a valuable locator of the star-forming sites \textit{within} the diffuse \ha{} emission region or HI extent of these objects \citep[e.g.,][]{Mundell:2004,Neff:2005}.

\subsection{Neutral and Molecular Hydrogen}
\objname{} is spatially coincident with the base of a several kpc-long structure known as the ``countertail" or Northern HI streamer \citep[][hereafter \yun{}]{Yun:1993}. This feature is theorized to have been torqued northward --- opposite another HI stream in the south-west which extends down toward M81 --- after the close passage of the two galaxies several hundred Myr ago. 

The overdensity of HI in the stream at the location of \objname{} can be seen in Figures \ref{M82_crop} and \ref{HI_comparison} (left). The HI gas mass for the full countertail, calculated via 
\begin{equation}
    M_{\rm HI} = 2.356\times10^5 D^2 \int S_\nu dv 
\end{equation}
where $S_\nu$ is in Jy and $D$ is in Mpc, is $\sim$2$\times$10$^{8}$ M$_{\odot}$, with $\sim$5.1$\times$10$^{7}$ M$_{\odot}$ in an aperture cospatial with the \ha{} source in the moment 0 map. For mass calculations we use the available zero-spacing corrected data cube from \deblok{} which combines the sensitivity of previous single dish observations \citep{Chynoweth:2008} with the high resolution data gathered by \deblok{}. The HI gas in this region has a mean velocity of 45 \kms{} with respect to M82's recession velocity, with a 1$\sigma$ spread of 47 \kms{} along the line of sight (Figure \ref{HI_comparison}, right). It is well known that this gas is kinematically decoupled from the disk of M82 \citep[e.g.,][]{Sofue:1998}, as the disk has a rotation velocity of $\sim$120 \kms{} at roughly the position of \objname{} \citep{Greco:2012}.

\begin{figure*}[tbh!]
    \centering
    \includegraphics[width=\linewidth]{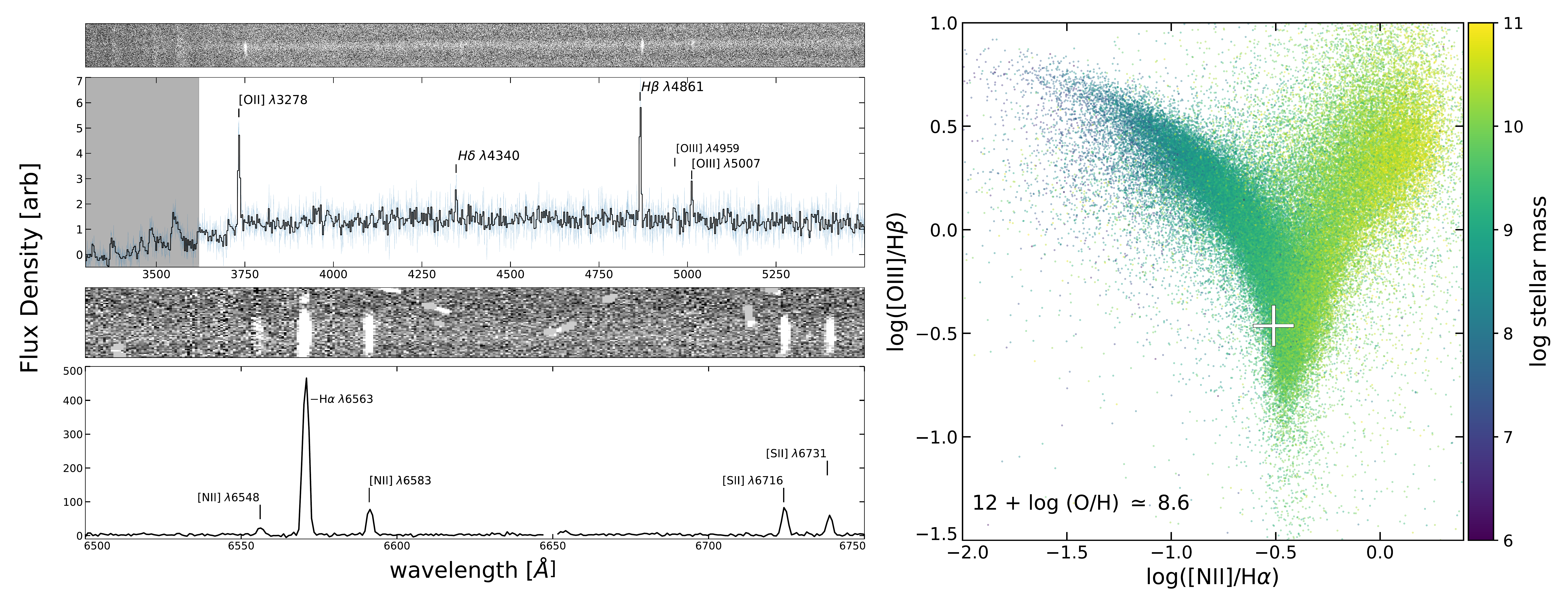}
    \caption{Left: Extracted LRIS spectrum of \objname{} in the blue (top) and red side (bottom). On the blue side, we show the raw spectrum in blue and one binned by 3 pixels in black. We readily detect \ha{}, both [NII] lines, both [SII] lines, H$\beta$, [OII], [OIII] (5007 \AA{} and 4959 \AA{}), and H$\delta$. The source's velocity is estimated by performing a Bayesian fit to four lines on the red side. Right: BPT diagram showing \objname{}'s location (white plus) relative to SDSS galaxies from the NASA Sloan Atlas, colored by stellar mass.}
    \label{spec1d}
\end{figure*}

Given the presence of star formation, we would also expect to see kinematically and spatially coincident molecular hydrogen in the general vicinity of \objname{}. Indeed, CO ($J=2\rightarrow 1$) data from \leroy{} show an overdensity in the 251 \kms{} channel. This corresponds to a velocity of 42 \kms{} with respect to M82's recession --- in strong agreement with the peak of HI emission (Figure \ref{HI_comparison}).

While the spatial alignment of \ha{} and FUV emission of \objname{} with that of HI and CO strongly suggests a star forming object within the tidal streamer, unambiguously confirming the association of the \ha{} region and ionizing sources with the tidal feature, rather than the disk of M82, requires detailed velocity information for \objname. To this end, we obtained optical spectroscopy with Keck.

\section{Keck / LRIS Spectroscopy}

Red side and blue side spectroscopy of \objname{} was carried out using the LRIS spectrograph \citep{Oke:1995,Rockosi:2010}, mounted on the Keck I telescope, on the nights of April 14, 2021 and May 07, 2021, respectively. While the primary goal was to obtain the velocity of the source, we also obtained relevant diagnostic line ratios, including [NII]/\ha{} and [OIII]/H$\beta$. 

Conditions during the runs were clear and the seeing was stable at $\sim$0.6$''$ and $\sim$1.3$''$ for the two nights, respectively. Exposures were obtained using the 1.5$''$-wide, 168$''$ longslit, using the 560 nm dichroic, 600/4000 grism, and 1200/7500 grating with a 39.6$^\circ$ dispersion angle to place 6563 \AA{} at the center of the red side detector. M82 was observed at its highest elevation, which corresponded to an airmass of $\sim$1.5. Two consecutive 600s exposures were obtained. The longslit was aligned with the brightest knot of emission determined from the \textit{HST} data with a position angle of $-15^\circ$.

Reduction was carried out using the \texttt{PypeIt} pipeline \citep{pypeit:josspub} using the settings for \texttt{lris\_red} and \texttt{lris\_blue}. Instrumental flats and arcs obtained at the beginning of the evening --- as well as skylines in the science exposures --- were used to trace the slit edges and perform wavelength calibration, sky modeling and sky subtraction.

Extracted 1-D spectra (observed frame) are presented in Figure \ref{spec1d}, with relevant emission lines denoted. Wavelengths are shown in vacuum, and a heliocentric correction of $-18$ \kms{} was applied during reduction. Due to a skyline near \ha{} which was not fully removed in the \texttt{pypeIt} reduction, we re-reduced the frames using sky backgrounds to measure the \ha{} flux.

To determine the velocity of \objname{}, we fit a model to the four other emission lines in the red using \texttt{emcee} \citep[][]{ForemanMackey:2013}. Emission lines were fit with Gaussians, with amplitudes and widths left free. We constrained the wavelength differences between the lines to their intrinsic separations, and fit for a single wavelength shift. 

We find a best-fit velocity of 276.7 \kms{}, with a 1$\sigma$ posterior spread of 1.12 \kms{}; the velocity uncertainty is dominated by the calibration uncertainty, which we estimate to be 25 \kms{}. Assuming a heliocentric-corrected velocity for M82 of 203 \kms{} \citep{devaucouleurs:1991}, this corresponds to a velocity of 73.7 \kms{} with respect to the recession velocity of M82, indicating conclusively that \objname{} is not embedded within the rotating disk ($\sim120$ \kms{}). This is shown in Figure \ref{HI_comparison} (right), in which \objname{}'s velocity is well within the envelope of the HI gas and well-separated from the rotation velocity of M82.

We also extract line ratios from our spectroscopy, finding [NII]/\ha{} $=0.31\pm0.015$ and [OIII]/H$\beta = 0.34\pm0.07$. Using the $O3N2$ metallicity calibration of \cite{Pettini:2004},
\begin{equation}
    12+\log(O/H) = 8.73 - 0.32 \times O3N2,
\end{equation}
we find that \objname{} has a metallicity of $8.67\pm0.14$, similar to metallicities found in other TDGs \citep[e.g.,][]{Lisenfeld:2008} and consistent with the star forming gas being pre-enriched --- as one would expect for gas liberated from the galactic disk of a massive spiral galaxy \citep[e.g.,][]{Duc:2000,Recchi:2015}. This is reflected in Figure \ref{spec1d} (right), where on the Baldwin, Phillips \& Telervich (BPT) diagram, \objname{} falls on the locus of star forming galaxies, here drawn from the NASA Sloan Atlas \citep[][]{Blanton:2011}, in a location consistent with M82's log stellar mass of $\sim$9-9.5.

\section{Discussion}
Based on these observations, \objname{} is an object similar to other claimed Tidal Dwarf Galaxies that have been recently discussed in the literature \citep[e.g.,][]{Mundell:2004,Querejeta:2021,Roman:2021}. If it survives the eventual dissipation of the tidal streamer and does not fall back into M82, it will join the ranks of dwarf satellite galaxies in the M81 group.  

Many objects of this type, however, are transient in nature, quickly falling back into the galactic disk \cite{Bournaud:2006}. The tidal tail which houses this complex of star forming regions is only several hundred Myr old, and numerical simulations suggest that a $\sim$Gyr of decoupled evolution is needed before one can unambiguously declare an object a TDG. 
\objname{} has properties that both aid and hinder its potential longevity. The velocity of the tidal arm and \objname{} is prograde with respect to M82's rotation, which has been found to be helpful in maintaining the TDG over time \citep{Bournaud:2006}. Conversely, the most successful TDGs seem to be those which form at the very tip of tidal arm structures (as far as possible from their original host). \objname{} is near the base of the tidal structure. An important factor governing its survival is the velocity at which the arm (and TDG progenitor within) is moving upward and away from M82; however, this motion is primarily in the sky-plane and is thus unconstrained.

Further complicating this picture is the broader environment of M82. Over the next Gyr, simulations indicate that the main galaxies in the group, M81, M82, and NGC 3077, are likely to merge \citep[][]{Yun:1999,Oehm:2017}, making the fate of this particular object an open question. It is interesting to consider whether any other TDGs may be forming in this region; if there are, they would likely be found farther up along the northern tidal streamer, or along the southern streamer toward M81. Deeper, higher S/N observations with the upgraded \delm{} will be needed to investigate this possibility.

Ultimately, \objname{} is a strong candidate for a young TDG, is in the earliest stages of its formation, and is remarkably nearby at $\sim$3.6 Mpc, making it a prime candidate for resolved studies of the earliest stages of this class of object.

\noindent\facility{ Keck:I (LRIS); Dragonfly Telephoto Array}

\vspace{10pt}
\software{\texttt{numpy} \citep{numpy:2011}, \texttt{astropy} \citep{astropy:2018}, \texttt{matplotlib} \citep{Hunter:2007}, \texttt{emcee} \citep{ForemanMackey:2013}, \texttt{PypeIt} \citep{pypeit:josspub}, \texttt{SExtractor} \citep{Bertin:1996}.}

\section*{Acknowledgements}
We thank the anonymous referee, whose comments and suggestions improved the quality of this work. We thank Todd Thompson and Natascha F\"{o}rster Schreiber for valuable discussion and feedback. I.P. is supported by the National Science Foundation Graduate
Research Fellowship Program under grant No. DGE1752134. Some of the data presented herein were obtained at the W. M. Keck Observatory, which is operated as a scientific partnership among the California Institute of Technology, the University of California, and the National Aeronautics and Space Administration. The Observatory was made possible by the generous financial support of the W. M. Keck Foundation. The authors wish to recognize and acknowledge the very significant cultural role and reverence that the summit of Maunakea has always had within the indigenous Hawaiian community.  We are most fortunate to have the opportunity to conduct observations from this mountain. This research is based on observations made with the NASA/ESA Hubble Space Telescope obtained from the Space Telescope Science Institute, which is operated by the Association of Universities for Research in Astronomy, Inc., under NASA contract NAS 5–26555. These observations are associated with program 10776 (PI: Matt Mountain). The authors thank the staff at New Mexico Skies Observatory for their support and aid in maintaining the \delm{} instrument. The Dunlap Institute is funded through an endowment established by the David Dunlap family and the University of Toronto. 

\bibliography{sample63}{}
\bibliographystyle{aasjournal}

\end{CJK*}
\end{document}